\documentstyle [12pt]{article}
\begin{document}
lgonzalz@vxcern.cern.ch
~~~~~~~~~~~~~~~~~~~~~~~~~~~~~~~~~~~~~~~~~~~~~~~~~~~~~~~~~~~~~~~~~~~~~~February
1997
\vskip 2.5cm
\centerline {\bf SPACE, TIME AND SUPERLUMINAL PARTICLES}
\vskip 2cm
\centerline {\bf L. GONZALEZ-MESTRES} 
\vskip 5mm
\centerline {Laboratoire de Physique Corpusculaire, Coll\`ege de France}
\centerline {11 pl. Marcellin-Berthelot, 75231 Paris Cedex 05 , France}
\vskip 3mm
\centerline {and}
\vskip 3mm
\centerline {Laboratoire d'Annecy-le-Vieux de Physique des Particules}
\centerline {B.P. 110 , 74941 Annecy-le-Vieux Cedex, France}
\vskip 3cm
{\bf Abstract}
\vskip 4mm
If textbook Lorentz invariance is actually 
a property of the equations describing a sector
of matter above some critical distance scale, 
several sectors of matter with different
critical speeds in vacuum can coexist and an absolute rest frame (the vacuum 
rest frame, possibly related to the local rest frame of the expanding Universe)
may exist without contradicting the apparent Lorentz invariance felt by 
"ordinary" particles (particles with critical speed in vacuum equal to $c$ ,
the speed of light). The real geometry of space-time will then be different
from standard Lorentz invariance, and the Poincar\'e relativity principle 
will be a local (in space and time), approximate sectorial property.
It seems natural to assume that particles with critical speed in vacuum
different from $c$ are superluminal.
\vskip 3mm
We illustrate such a scenario using as an example a spinorial space-time where
the modulus of the spinor, associated to the time variable, is the size of 
an expanding Universe. Several properties of superluminal particles,
and of matter without a universal relativity principle, are
discussed in view of experimental applications. If the vacuum rest frame is
close to that suggested by the cosmic microwave background, experimental
searches for superluminal particles on earth should mainly contemplate a
laboratory speed range around $10^3~c$ , even for very high energy 
superluminal cosmic rays. The detectability of several consequences of
the new scenario is briefly discussed.
\vskip 20cm
{\bf 1. INTRODUCTION} 
\vskip 5mm
In recent papers [1-5] , we pointed out that the apparent Lorentz invariance of
the laws of physics does not imply that space-time is indeed minkowskian. 
Lorentz invariance can be just a property of the equations describing a sector
of matter above some critical distance scale. 
Then, an absolute rest frame (possibly related to the local rest frame of the
expanding Universe) may exist without contradicting the apparent Lorentz
invariance felt by the particles we are made of. We 
suggested the possible existence of superluminal sectors of matter, i.e. of
particles with positive mass and energy but with a critical speed in vacuum
much higher than the speed of light $c$~ . Sectors of matter with different
critical speeds in vacuum may coexist, just as in a perfectly transparent
crystal it is possible to identify at least two critical speeds: those of sound
and light. Interaction between the "ordinary" sector, i.e. particles with
critical speed in vacuum equal to $c$ , and the superluminal sectors, would 
break Lorentz invariance. But such interactions would be expected to be
basically very high-energy, short distance phenomena not incompatible with 
successful conventional tets of Lorentz invariance.
Several important physical and 
cosmological implications were discussed in a scenario with several critical
speeds for particles in vacuum. 
\vskip 4mm
Admitting the possible
existence of several sectors of matter with different critical
speeds in vacuum, some stringent form of grand-unified or primordial constituent
background seems necessary in order to explain why so
many different particles
(quarks, leptons, gauge bosons...) have the same critical speed. This is
in agreement with present ideas in particle theory,
and to some extent improves their formulation as it appears that
grand unification and universality of the critical speed in vacuum
may be expressions of a single 
symmetry inside each sector. Our hypothesis extends
the current approach, but does not really contradict its basic philosophy.
However, the superluminal particles we propose are definitely not
space-like states of the ordinary ones: they are a new kind of matter related
to new degrees of freedom not yet discovered experimentally.
If superluminal particles exist, they may considerably modify the Big Bang
scenario, 
provide most of the cosmic (dark) matter, lead the evolution of the
Universe, influence low-energy physics and be
produced at very high-energy accelerators (e.g. LHC) 
or found in experiments devoted to
high-energy cosmic rays (e.g. AMANDA [6]) where they can yield very specific
signatures allowing to detect events at extremely small rates (e.g. [5] ,
but see also Sect. 4). 
\vskip 4mm
In this note, we would like to discuss some possible implications of the new
approach for our description of space, time
and elementary particles. As an example, a previously
considered $SU(2)$
spinorial space-time [4 , 5] will be chosen as the framework. If the structure
of space-time reflects basically the properties of matter at the scales under
consideration, it should to some extent account for the structure and evolution
of the Universe, as well as for phenomena like spin 1/2 which cannot be 
described in a natural way using conventional space-time coordinates. 
In general relativity,
the gravitational properties of matter modify the local metric of space-time,
but a much closer connection between matter and space-time can be imagined
incorporating deeper dynamical properties. Also, departure from the Poincar\'e 
relativity principle
yields new fundamental physics (new particles and interactions,
motion "backward in time" in some frames...), as will be discussed later. 
\vskip 6mm 
\vskip 6mm
{\bf 2. THE SPINORIAL SPACE-TIME}
\vskip 5mm
Instead of four real numbers, we take space-time to be described by two complex 
numbers, the components of a $SU(2)$ spinor. From a spinor $\xi $ , it is
possible to extract a $SU(2)$ scalar, 
$\mid \xi \mid ^2$ $=$ $\xi ^\dagger \xi $ (where the dagger stands for
hermitic conjugate), and a 
vector ${\vec {\mathbf z}}~=~\xi ^\dagger {\vec \sigma \xi }$ , where ${\vec
\sigma }$ is the vector formed by the Pauli matrices. In our
previous papers on the subject [4 , 5] , we proposed to interpret $t$ = 
$\mid \xi \mid $ as the time. 
If the spinor coordinates are complex numbers, one has: $z$ = $t^2$ 
where $z$ is the modulus of ${\vec {\mathbf z}}$ .
It does not seem possible to interpret ${\vec {\mathbf z}}$ 
as providing the space
coordinates: one coordinate, corresponding to an
overall phase of the spinor, is missed by $t$ and ${\vec {\mathbf z}}$ . 
Therefore,
a different description of space seems necessary in this approach.
\vskip 4mm
Interpreting $t$ as the time has at first sight the drawback of
positive-definiteness and breaking of time reversal, 
but this can be turned into an advantage if $t$ is
interpreted as an absolute, cosmic time (geometrically expanding Universe).
An arrow of time is then naturally set, and space-time geometry incorporates 
the physical phenomenon of an expanding Universe. 
As space translations and rotations are by definition transformations leaving
time invariant, the space coordinates should be
built by considering the polar coordinates in the hypersphere (i.e. the 
spherical hypersurface) of constant
time, $\mid \xi \mid $ = $t_0$ where $t_0$ is a value of time [4 , 5]. 
On this hypersphere, a point $\xi $ can be decribed as:  
\equation
\xi ~=~ U\xi _0
\endequation
where $U$ is a $SU(2)$ transformation and $\xi _0$ a constant spinor (hereafter
identified with the observer position) on the
sphere $t$ = $t_0$ . Writing:
\equation
U~=~exp~(i/2~~t_0^{-1}~{\vec \sigma }.{\vec {\mathbf x}})~
\equiv U({\vec {\mathbf x}}) 
\endequation
the vector ${\vec {\mathbf x}}$ , with $0~\leq x$ 
(modulus of ${\vec {\mathbf x}}$) 
$\leq$  $2\pi t_0$ , can be interpreted 
as the position vector at constant
time $t_0$ . It is unique, except for
a $2\pi $ rotation ($x~=~2\pi t_0~ , U~=~-1$). The natural metric is
$dr^2({\vec {\mathbf x}}~,~ 
{\vec {\mathbf dx}})~=~({\vec {\mathbf dx^*}})^2$ ,               
where ${\vec {\mathbf dx^*}}$
is defined by:
\equation
U({\vec {\mathbf x}}~+~ {\vec {\mathbf dx}})~~=~~U({\vec {\mathbf dx^*}})~
U({\vec {\mathbf x}})
\endequation
or, in terms of exponentials:
\equation
exp~[i/2~~t_0^{-1}~{\vec \sigma }.({\vec {\mathbf x}}+{
\vec {\mathbf dx}})]~~=
~~exp~(i/2~~t_0^{-1}~{\vec \sigma }.{\vec {\mathbf dx^*}})~
exp~(i/2~~t_0^{-1}~{\vec \sigma }.{\vec {\mathbf x}})
\endequation
leading at infinitesimal level to:
\equation
{\vec {\mathbf dx^*}}~~~=~~~{\vec {\mathbf dx_L}}~~+~~
2t_0~[cos~(t_0^{-1}x/2)~sin~(t_0^{-1}x/2)~x^{-1}~ 
{\vec {\mathbf dx_{\perp }}}~~-~~sin^2~(t_0^{-1}x/2)~x^{-2}
~{\vec {\mathbf x}}~
\wedge ~{\vec {\mathbf dx}}]
\endequation
where ${\vec {\mathbf dx_L}}~=~x^{-2}~
({\vec {\mathbf x}}.{\vec {\mathbf dx}})~{\vec {\mathbf x}}$
and ${\vec {\mathbf dx_{\perp }}}~=~ {\vec {\mathbf dx}}-
{\vec {\mathbf dx_L}}$ . From (5), 
we consistently recover ${\vec {\mathbf dx^*}}
(x=0)~=~ {\vec {\mathbf dx}}$ ~, whereas  ${\vec {\mathbf dx^*}}
(x=2\pi t_0)~=~ {\vec {\mathbf dx_L}}$ . 
Any closed path of the form:
$U(\alpha {\vec {\mathbf u}})~=
~exp~(i/2~~\alpha t_0^{-1}~{\vec \sigma }.{\vec {\mathbf u}})$
from $\alpha =-2\pi t_0$ to $\alpha =2\pi t_0$ , 
where ${\vec {\mathbf u}}$ is a unitary 
three-dimensional real vector, defines a geodesic on the hypersphere and
on  $SU(2)$ . The local volume element at $\xi $ is $d^3{\vec {\mathbf x^*}}~
=~dx^*_1dx^*_2dx^*_3$ .
\vskip 4mm
It 
is obvious that, under a $SU(2)$ transformation $V$ , $U$ transforms 
into $VUV^{-1}$ (the vector representation) if 
${\vec \sigma }.{\vec {\mathbf x}}$  
transforms into $V~{\vec \sigma}.{\vec {\mathbf x}}~V^{-1}$ 
(the vector linear representation). 
The vector ${\vec {\mathbf v}}$
obtained from the equation $V~=~exp~(i/2~~{\vec \sigma }.{\vec {\mathbf v}})$ 
defines the rotation axis and angle in correspondence with $SO(3)$
rotations. ${\vec {\mathbf x}}$ provides the space coordinates and
transforms like a $SO(3)$ real vector,
but being a $SU(2)$ parameter it varies on a spherical volume of radius 
$2\pi t_0$ , $x~\leq ~2\pi t_0$ , 
whose surface is identified to a single point.
Under the rotation defined by $V$ , ${\vec {\mathbf x}}$ transforms into $R_V
{\vec {\mathbf x}}~~=~~{\vec {\mathbf x}}_L~+
~cos~v~{\vec {\mathbf x}}_{\perp }~
+~v^{-1}~sin~v~{\vec {\mathbf x}}~
\wedge ~{\vec {\mathbf v}}$~, where ${\vec {\mathbf x}}_L~~=~~
v^{-2}~({\vec {\mathbf x}}.{\vec {\mathbf v}})~{\vec {\mathbf v}}$ and
${\vec {\mathbf x_{\perp }}}~=~ {\vec {\mathbf x}}-{\vec {\mathbf x}}_L$ ,
as it should be the case for a genuine space rotation. The observer can 
thus feel a vector space in a spinorial space-time.
\vskip 4mm
The cosmic time scale given by the radius of the Universe 
does not correspond to the local time scale of 
physical processes at time $t$ , which depends on the local vacuum dynamics
and varies with cosmic time. In situations where the Universe evolves smoothly,
we can identify both time scales up to a constant to be determined locally.
Whatever the relation between the two time scales, the above example 
(which requires by itself unification of space and time) clearly 
shows that abandoning textbook relativity dogmas does not necessarily 
destroy the standard cosmological framework.
\vskip 6mm
{\bf 3. SPACE-TIME SYMMETRIES AND WAVE FUNCTIONS}
\vskip 5mm
On a three-dimensional vector representation, a $2\pi $ rotation is equal to
the
identity, contrary to spinor representations of $SU(2)$ where the same
transformation changes the sign of the spinors. The
description of particles 
with half-integer spin in a real space-time has always posed
some conceptual problems, in the sense that a spinor wave function cannot be
built by standard operations from representations of the space rotation group.
More precisely, a particle with half-integer spin cannot be described by
a single-valued function of the standard space and time variables. 
To circumvent this difficulty,  
it is  said that spinors do not form representations of the basic symmetry
group $SO(3)$ of space but of the covering group $SU(2)$ 
of its Lie algebra, or that they belong to representations of the rotation
group "up to a sign". The same situation arises in a
relativistic description, where 
$SL(2~,~C)$ is the covering group of the Lie algebra
of the Lorentz group which, when complexified, is equivalent to that of
$SL(2~,~C)_{left}~\otimes ~SL(2~,C)_{right}$ (for "left" and "right" chiral
spaces).
Our spinorial space-time naturally provides a more compact description:
position in the spinor space-time is described by a spinor whereas,
simultaneously and independently of whether relativity applies, 
position in the $\mid \xi \mid  ~=~t_0$ hypersphere 
(the "space" at time $t$ = $t_0$) is
described by a vector. Spinor fields become then single-valued functions of 
space-time coordinates. It seems worth emphasizing again 
that, even without relativity,
our approach requires and operates a geometric unification of 
space and time. 
\vskip 4mm
At fixed $t_0$ , the position vector ${\vec {\mathbf x}}$  
describes the relative position
of the point $\xi $ with respect to an observer placed at $\xi _0$ .
The space coordinates are the parameters of the (unique) element $U$ of $SU(2)$
which transforms $\xi _0$ into $\xi $ . 
Contrary to the conventional description of 
space rotations, no $SU(2)$ transformation
other than the identity leaves the observer position unchanged, but the
observer can measure only the relative position of surrounding objects.
A $2\pi $ rotation would 
change the signs of $\xi _0$ and $\xi $ simultaneously. It forms 
with the identity the center of the group and cannot be felt by any
position measurement (the observer is by definition insensitive to its own
motion), 
although the properties of
spin-1/2 particles can be checked by other experiments. Therefore, 
in our approach, a
space rotation around an axis passing through the observer position is 
a $SU(2)$ transformation whose effects cannot be entirely seen through 
position measurements. If $V$ is the $SU(2)$ transformation, all the points 
in the spinor space of
the form $\xi $~ = ~$V^{\lambda }~\xi _0$ , where $\lambda $ is a real number
between $- 2\pi v^{-1}$ and $2\pi v^{-1}$ , 
appear unchanged to the observer when $V$ acts on the spinor
space-time. However, $V$ actually changes all these points (they move along the
rotation axis)
and preserves only the
overall geodesic on the $\xi $ = $t_0$ hypersphere. 
Space
rotations significantly transform the whole Universe, and can deplace
the observer at cosmic scales.
While space rotations correspond to $SU(2)$ transformations, space
translations correspond to a change in the position of the observer on the 
$\xi $ = $t_0$ hypersphere. Moving the observer from $\xi _0$ to $\xi _0'$ =
$W\xi _0$ changes the above defined 
$U$ into $UW^{-1}$ and, writing 
$W~=~exp~(-i/2~~t_0^{-1}~{\vec \sigma}.{\vec {\mathbf w}})$ , 
${\vec \sigma}.{\vec {\mathbf x}}$ is changed into $-2it_0~ln~(UW^{-1})$ ,
where $ln$
stands for neperian logarithm uniquely defined, for $U~\neq ~-W$ , in the 
range of traceless hermitic $2 \times 2$ matrices with eigenvalues 
$\pm \lambda $ such that $\mid \lambda \mid ~<~2\pi t_0$ .    
In the "local" limit $x/t_0~\ll ~1$ and $w/t_0 ~\ll ~1$ , 
the expression  $-2it_0~ln~(UW^{-1})$ can 
be approximated by
${\vec \sigma }.({\vec {\mathbf x}}~+~{\vec {\mathbf w}})$.
Space translations commute in the infinitesimal limit where 
the structure of the Lie algebra (providing the cosmic 
curvature) can be ignored.
The usual parametrization of space
transformations is a local one, based on our intuitive, infinitesimal 
view of the tangent
hyperplane to the $\xi $ = $t_0$ hypersphere at the observer position. 
\vskip 4mm
Contrary to space translations, space-time translations defined using spinorial
coordinates are real translations which commute but which cannot leave
invariant any $\mid \xi \mid $ = constant hypersphere. We expect them to make
sense locally, at scales much smaller than the size of the Universe.
The generators of such
translations are $SU(2)$ spinors, and so is the space-time 
position spinor $\Delta \xi
$ = $\xi -\xi _0$ of any point of space-time with respect 
to an observer placed at $\xi _0$ . Supersymmetry could possibly be connected 
to derivation with respect to the spinorial coordinates, but we shall not
discuss this point in the present work.
\vskip 4mm
Because of the expansion of the space
hypersphere, we do not really expect invariance under time translations
(energy conservation) to hold at cosmic scale, even if it can be a good local
symmetry in the infinitesimal limit $\Delta t~\ll ~t$ where our intuition 
suggests energy conservation to be a basic law of Nature. Similar
considerations apply to discrete time symmetries, such as $T$ and $CPT$
(no longer protected by a universal Lorentz invariance),
as the two directions of time are clearly inequivalent at cosmic scale.
As will be seen later, we expect the cosmological redshift and time dilation
to apply as in standard cosmology.
\vskip 4mm
Up to a $n$-dependent normalization factor, and taking the above 
$d^3{\vec {\mathbf x^*}}$ as the volume element,
an example of spinor wave function at time $t$ could be, 
for integer $n~\geq 0$ :
\equation
\psi _n(\xi  )~~=
~~t^{-3/2}~(\psi _0^{\dagger }~U~  \psi _0)^n~\psi _0
\endequation
and, for integer $n~\leq 0$ :
\equation
\psi _{-n}~(\xi )~~=
~~t^{-3/2}~(\psi _0^{\dagger }~U^{\dagger }~\psi _0)^n~\psi _0
\endequation
where $\xi _0 $ is the observer position at time $t$ ,
$\psi _0$~ =~ $\psi $ ($\xi _0$) is the
value of the spinor wave function at $\xi _0$ and $U$ is the operator uniquely
defined by equation (1) , tranforming $\xi _0$ into $\xi $ . The quantization 
of $n$ is required by the regularity of the wave function at zero values
of the matrix elements $\psi _0^{\dagger}~U~ \psi _0 $ . In 
realistic quantum-mechanical situations, $n$ is a
very large number in order to make the wave function oscillate at small
wavelengths. 
The function defined by the $n$-th power of the 
$n~=~1$ wave function $\psi _0^{\dagger}~U~\psi _0$ , or by its complex
conjugate, 
is then close to a plane wave, as can be seen in what follows. 
\vskip 4mm
From the parametrization (1) , we can replace $\xi $ by its space coordinate 
vector ${\vec {\mathbf x}}$ and 
write for a spinor wave function $\psi ~~=~~\psi~({\vec {\mathbf x}}~,~t)$. 
Let us the consider the case where $\psi _0$ is an
eigenspinor of ${\vec {\sigma }}.{\vec {\mathbf x}}$ , 
and set a coordinate system 
where ${\vec {\mathbf x}}$ = $(x~ ,~ 0~ ,~ 0)$ . 
Then, if $\sigma _1$ is taken to be diagonal and 
$\sigma _1~\psi _0~=~\psi _0$ , one has 
$\psi _0^{\dagger}~U~\psi _0~~=~~exp~(it^{-1}x/2)$ . If, instead, 
${\vec {\mathbf x}}~~=~~(x_1~ ,~ x_2~ ,~ x_3)$ ,
and if $x_2~, x_3 ~\ll ~t $ , one has 
~$\psi_0^{\dagger} ~U~\psi _0~~\simeq~~exp~(it^{-1}x_1/2)$. In both cases, the 
wave function can be approximated by a plane wave
with wave vector ($t^{-1}/2$ , 0 , 0).
Similarly, the $n$-th power of the matrix element will be a plane wave with
momentum ($nt^{-1}$/2 , 0 , 0) and can be approximated by
$exp~(i~n~t^{-1}~x_1/2)$ as long as
$~n~(x^2~-~x_1^2)~~\ll ~~t^2$ . The above considered 
translation $W$ , acting on the spinor wave functions, 
will turn $U$ into $UW^{-1}$ . It will, to leading order, transform 
${\vec {\mathbf x}}$ into ${\vec {\mathbf x}}~+~{\vec {\mathbf w}}$
and the wave function $\psi _n~ \propto
~exp~(i~n~t^{-1}~x_1/2)$ into 
$\psi '_n~\propto ~exp~[i~n~t^{-1}~(x_1~+~w_1)/2]$ as long as 
~$n~ (w^2~-~w_1^2)~~\ll ~~t^2$ . 
On an arbitrary spinor wave function $\psi _n$ (and similarly for $\psi _{-n}$
via complex conjugation), 
of the form (6) , the generators of space translations
are the matrices $in/2~\sigma_j~t^{-1}$ ($j~=~1~,~2~,~3)$ 
multiplying $U$ at right inside the matrix element 
$\psi _0^{\dagger }U\psi_0 $ . Taking 
$t~\approx ~10^{26}~m$ for cosmic time,  
the above wave function can be approximated by a plane wave for
$x_\perp ^2~=~x^2~-~x_1^2~  
\stackrel{<}{\sim } ~10^{26}~m~ k_L^{-1}$ where $k_L~=~nt^{-1}/2$ 
is the longitudinal wave vector 
scale. If $n~\approx 10^{45}$ and 
$k_L^{-1}~\approx 10^{-19}~m$ (the lowest distance scale  
accessible to experiments at $TeV$ energies), the plane wave 
approximation holds for $\mid x_\perp \mid ~\stackrel{<}{\sim } ~ 1~Km $ 
(gaussian damping).
If $n~\approx 10^{61}$ and 
$k_L^{-1}~\approx 10^{-35}~m$ , the approximation is valid for
$\mid x_\perp \mid ~\stackrel{<}{\sim } ~ 10^{-5}~m$ .
Assuming that this transverse damping of quantum wave functions 
at high wavelength is
to be taken seriously, it is not clear whether feasible experiments can 
eventually measure such an effect related to the possible
finite radius of
the Universe.
\vskip 4mm
If $\psi _0$ is an eigenspinor of
$\sigma _1$ with eigenvalue $s $ (necessarily 1 or -1), 
and with the same approximations as before, $\psi _n$ will be a momentum
eigenstate with wave vector ${\vec {\mathbf k}}$ of components
($nst^{-1}$/2 , 0 , 0). It 
will therefore, with the same approximations,
satisfy the equation:
\equation
{\vec {\sigma }}.{\vec {\mathbf k}}~\mid \psi _n>~~=~~s ~k~n~\mid
n \mid ^{-1}~\mid \psi >~~=~~\lambda ~k~\mid \psi _n>
\endequation
where $\lambda ~=~s~n\mid n \mid ^{-1}$ is the helicity, with n taking 
integer values from (6) and (7).
\vskip 6mm
{\bf 4. EQUATIONS OF MOTION}
\vskip 5mm
In the present geometric approach, a natural cosmic time is given by the 
size of the 
Universe.
If radial directions from the point $\xi = 0$ define the arrow of time and 
the motion of an absolute local rest frame (the "vacuum rest frame"), 
it appears from 
continuity and from quantization 
due to the compactness of three-dimensional space, e.g. in (6) ,
that the wave 
vector ${\vec {\mathbf k}}$ of a quantum particle
must evolve in time like $t^{-1}$ , generating a
geometric shift of wavelengths due to the expansion of the Universe. 
This is not surprising, as it just corresponds to the geometric 
version of the standard cosmological redshift. 
Since our cosmic time $t$ is identical to
the radius of the Universe, we immediately recover
the well-known expression (e.g. [7]):
\equation
\lambda _o \lambda _e^{-1}~~=~~a_oa_e^{-1}
\endequation  
where $\lambda _o$ is the observed wavelength of the particle at the time
of its detection, $a_o$ the expansion parameter of the Universe at detection
time, $\lambda _e$ the emission wavelength and $a_e$ the value of the
expansion parameter at the time of emission of the particle. 
The recession rate, given by Hubble's constant, is very small
as compared to laboratory time scales. 
In the rest of the chapter, we will restrict ourselves to
local phenomena and ignore effects related: a) to time evolution 
at cosmic scale; b) to the curvature of space at large scales. In 
a global description, the speed ${\vec {\mathbf v}}$ = $d{\vec {\mathbf x}}/dt$
should be replaced by ${\vec {\mathbf v^*}}$ = ${\vec {\mathbf dx^*}}/dt$ ,
and the gradient with respect to ${\vec {\mathbf dx}}$ 
by a gradient with respect to ${\vec {\mathbf dx^*}}$ . Unless otherwise 
stated, we consider in what follows
the limit where the Universe is very large as compared 
to local space and time scales for laboratory phenomena,
and therefore we approximate
the wave vector spectrum by a continuum spectrum. 
\vskip 5mm
{\bf 4a. Classical mechanics and kinematics}
\vskip 5mm
To illustrate classical mechanics with superluminal particles, 
assume a system of N interacting
particles in the vacuum rest frame with: a) different critical speeds
$c_1~,~c_2~,~...~,~ c_N$ 
associated to different sectorial Lorentz invariances;
b) the following lagrangian in the local rest frame
of vacuum:
\equation
L~~=~~-~\Sigma _{i=1}^N~m_i~c_i^2~(1~-~v_i^2c_i^{-2})^{1/2}~~-
~~U({\vec {\mathbf x_1}}~,~{\vec {\mathbf x_2}}~,~...~,~{\vec {\mathbf x_N}})
\endequation
where: a)  $v_i$ is the modulus of ${\vec {\mathbf v_i}}$ ; b) the vectors
${\vec {\mathbf v_i}}$
and ${\vec {\mathbf x_i}}$ stand for the speed and position of
particle $i$ ; c) $m_i$ is the inertial mass of particle $i$ ; 
d) $U$ is a potential energy describing the interaction,
where we ignore large scale geometric effects related to the space coordinates.
This lagrangian implies that all sectorial Lorentz invariances
(i.e. all sectorial Lorentz metrics in their canonical diagonal form)
can be simultaneously exhibited in a single rest frame,
the "absolute" or "vacuum" rest frame. More complicate scenarios can be 
imagined.
Standard use of the variational principle 
leads from (10) to the equations of motion
$(i~=~1~,~2~,~...~,~N)$:
\equation
d/dt ~{\vec {\mathbf p_i}}~~=~~- {\vec {\mathbf \nabla _i}}~U
\endequation
where ${\vec {\mathbf \nabla _i}}$ means partial gradient with respect to
the position of the $i$-th particle and:
\equation
{\vec {\mathbf p_i}} =
m~ {\vec {\mathbf v_i}}~(1~-~v_i^2c_i^{-2})^{-1/2}
\endequation
\par
As in standard mechanics, momentum conservation holds:
\equation
d/dt~ (\Sigma  _{i=1}^N ~ {\vec {\mathbf p_i}})~~=~~0
\endequation
and energy conservation is expressed by the formula:
\equation
d/dt~ (\Sigma  _{i=1}^N ~ {\vec {\mathbf p_i}}.{\vec {\mathbf v_i}}~~-~~L~~)=
~~d/dt~ (\Sigma  _{i=1}^N ~ p^0_i c_i ~~+~~U)~~=~~0
\endequation
where:
\equation
p^0_i~~=~~E_ic_i^{-1}~~=~~m_ic_i~(1~-~v_i^2c_i^{-2})^{-1/2}
\endequation
and $E_i$ is the non-interacting (i.e. rest + kinetic) energy of particle $i$ .
${\mathcal{E}} ~~=~~\Sigma  _{i=1}^N ~ p^0_i c_i ~~+~~U$ is the total energy 
of the system.
The 4-vectors $(p^0_i~,~ {\vec {\mathbf p_i}})$ transform covariantly under
Lorentz tranformations with critical speed parameter $c_i$ .
However, the conservation laws involving several sectors with different 
values of the $c_i$'s are not covariant and
therefore can be written in the above form only in the vacuum rest frame.
"Lorentz" transformations to other inertial frames will depend on 
the matter the observer is made of. Since we expect to measure the energy 
of superluminal particles through interactions with "ordinary" particles,
we can define, in the rest frame of an "ordinary" particle moving at speed
${\vec {\mathbf V}}$ with respect to the vacuum rest frame, the energy and 
momentum of a superluminal particle to be the Lorentz-tranformed 
of its vacuum rest frame energy and 
momentum taking $c$ as the critical speed parameter for the Lorentz
transformation. Then, the mass of the superluminal particle will depend
on the inertial frame. The energy and momentum of particle $i$ in the new rest
frame, as measured by ordinary matter from energy and momentum 
conservation (e.g. in decays of superluminal particles into ordinary
ones), will be:
\equation
E'_i~~=~~(E_i~-~{\vec {\mathbf V}}.{\vec {\mathbf p_i}})~
(1~-~V^2c^{-2})^{-1/2}
\endequation
\equation
{\vec {\mathbf p'_i}}~~=~~{\vec {\mathbf p'_i}}_{\mathbf {,L}}~+~
{\vec {\mathbf p'_i}}_{\mathbf {,\perp }}
\endequation
\equation
{\vec {\mathbf p'_i}}_{\mathbf {,L}}~~=~~({\vec {\mathbf p_i}}_{\mathbf {,L}}~
-~E_i~c^{-2}~{\vec {\mathbf V}})~(1~-~V^2c^{-2})^{-1/2}
\endequation
\equation
{\vec {\mathbf p'_i}}_{\mathbf {,\perp }}~~=~~
{\vec {\mathbf p_i}}_{\mathbf {,\perp }}
\endequation
where  ${\vec {\mathbf p_i}}_{\mathbf
{,L }}~=~V^{-2}~
({\vec {\mathbf V}}.{\vec {\mathbf p_i}})~{\vec {\mathbf V}}$~ ,~
${\vec {\mathbf p_i}}_{\mathbf {,\perp }}~~=~~
{\vec {\mathbf p_i}}~-~{\vec {\mathbf
p_i}}_{\mathbf {,L }}$
and similarly for the longitudinal and transverse components 
of ${\vec {\mathbf p'_i}}$ .
We are thus led to consider the effective squared mass:
\equation
M_{i,c}^2~~=~~c^{-4}~(E_i^2~-~c^2p_i^2)~~=~~
m_i^2c^{-4}c_i^4~+~c^{-2}(c^{-2}c_i^2~-~1)~p_i^2
\endequation
which depends on the vacuum rest frame momentum of the particle.
$m_i$ is the invariant mass of particle $i$ , as seen by matter from the 
$i$-th 
superluminal sector (i.e. with critical speed in vacuum = $c_i$).
While "ordinary" transformation laws of energy and momentum are not singular,
even for a superluminal particle, the situation is different for  
the transformation of a superluminal
speed, as will be seen below. Furthermore,
a mathematical surprise arises: assume ${\vec {\mathbf v_i}}~=~
{\vec {\mathbf V}}$ , so that particle $i$ is at rest in the new 
inertial rest frame. Then, we would naively expect a vanishing momentum,
${\vec {\mathbf p'_i}}~=~ {\mathbf 0}$ . Instead, we get: 
\equation
{\vec {\mathbf p'_i}}~~=
~~-~{\vec {\mathbf p_i}}~(c^{-2}c_i^2~-~1)~(1~-~V^2c^{-2})^{-1/2}
\endequation
and $p'_i~\gg ~p_i$ , although $p'_ic~\ll ~E'_i$ if $V~\ll ~c$ .
This reflects the non-covariant character of the 4-momentum of particle $i$
under "ordinary" Lorentz transformations. Thus, even if the directional
effect is small 
in realistic situations (f.i. on earth), 
the decay of a superluminal particle at rest into 
ordinary particles will not lead to an exactly vanishing total momentum if the
inertial frame is different from the vacuum rest frame.
\vskip 4mm
In the rest frame of an "ordinary" particle moving with speed 
${\vec {\mathbf V}}$
with respect to the vacuum rest frame, we can estimate the speed 
${\vec {\mathbf
v'_i}}$ of the previous particle $i$ writing:
\equation
{\vec {\mathbf v_i}}~~=~~{\vec 
{\mathbf v_i}}_{\mathbf {,L}}~+~{\vec {\mathbf v_i}}_{\mathbf{,\perp }}
\endequation
where ${\vec {\mathbf v_i}}_{\mathbf 
{,L }}= V^{-2}
({\vec {\mathbf V}}.{\vec {\mathbf v_i}})~{\vec {\mathbf V}}$~ ,~
${\vec {\mathbf v_i}}_{\mathbf {,\perp }}=
{\vec {\mathbf v_i}}-{\vec {\mathbf
v_i}}_{\mathbf {,L }}$ 
and similarly for the longitudinal and transverse components 
of ${\vec {\mathbf v'_i}}$ . Then, the transformation law is:
\equation
{\vec {\mathbf v'_i}}_{\mathbf {,L}}~~=~~({\vec {\mathbf v_i}}_{\mathbf {,L}}~
-~{\vec {\mathbf V}})~
(1~-~{\vec {\mathbf v_i}}.{\vec {\mathbf V}}~ c^{-2})^{-1}
\endequation
\equation
{\vec {\mathbf v'_i}}_{\mathbf {,\perp }}~~=~~{\vec {\mathbf v_i}}_{\mathbf 
{,\perp}}~(1~-~V^2c^{-2})^{1/2}~(1~-~
{\vec {\mathbf v_i}}. {\vec {\mathbf V}}~c^{-2})^{-1}
\endequation 
leading to singularities at ${\vec {\mathbf v_i}}.{\vec {\mathbf V}}~~=~~c^2$ 
which correspond to a change in the arrow of time (due to the Lorentz 
transformation of space and time with respect to the 
"absolute" rest frame) as seen by ordinary matter
traveling at speed ${\vec {\mathbf V}}$ with respect to the vacuum rest frame.
At $v_{i,L}~>~c^2V^{-1}$ , a superluminal particle moving forward in time in
the vacuum rest frame will appear as moving backward in time to
an observer made of ordinary matter and moving at speed ${\vec {\mathbf V}}$
in the same frame. On earth, taking $V~\approx ~10^{-3}~c$ 
(if the vacuum rest frame is close to that suggested by cosmic background
radiation, e.g. [7]), the apparent
reversal of the time arrow will occur mainly at 
$v_i~\approx ~ 10^{3}~c~.$ If $c_i~\gg ~10^3~c$ ,
phenomena related to propagation backward in time of produced superluminal
particles may be observable in future
accelerator experiments slightly above the
production threshold. In a typical event where a pair of superluminal particles
would be produced, we expect in most cases that one of the superluminal
particles propagates forward in time and the other one propagates backward. 
It must be noticed that, according to (16~-~19), the infinite velocity (value
of $v'_i$)
associated to the point of time reversal does not correspond to infinite
values of energy and momentum.
The backward propagation in time, as observed by devices which are not at
rest in the vacuum rest frame, is not really physical 
(the arrow of time is well defined in the vacuum rest frame for all
physical processes) and does not 
correspond to any real violation of causality (see also Subsection 4d). 
The apparent reversal of the time arrow for superluminal particles at
${\vec {\mathbf v_i}}.{\vec {\mathbf V}}~>~c^2$ would 
be a consequence of 
the bias of the laboratory time measurement due to our motion
with respect to the absolute rest frame. 
The distribution and properties
of such events, in an accelerator experiment or in a large volume cosmic ray
detector,
would obviously be in correlation with the direction and 
speed of the laboratory's motion with respect to the absolute rest frame
and provide fundamental cosmological information, complementary to cosmic
microwave background.
\vskip 4mm
From (23) and (24), we also notice that, for $V~\ll ~c$ and 
${\vec {\mathbf v_i}}. {\vec {\mathbf V}}~\gg ~c^2$ , the speed 
${\vec {\mathbf v'_i}}$ tends to the limit 
${\vec {\mathbf v_i}}^{\infty }$ , where:
\equation
{\vec {\mathbf v_i}}^{\infty }~
({\vec {\mathbf v_i}})~~=~~-~{\vec {\mathbf v_i}}~c^2~
({\vec {\mathbf v_i}}.{\vec {\mathbf V}})^{-1}
\endequation
which sets a universal high-energy limit, independent of $c_i$ , 
to the speed of superluminal particles as
measured by ordinary matter in an inertial rest frame other than the vacuum
rest frame. This limit is not isotropic, and depends on the angle between 
the speeds ${\vec {\mathbf v_i}}$ and ${\vec {\mathbf V}}$ . 
A typical order of magnitude for  
${\vec {\mathbf v_i}}^{\infty }$ on earth is
${\vec {\mathbf v_i}}^{\infty }~\approx 10^3~c$ if 
the vacuum rest frame is close to
that suggested by cosmic background radiation.
If $C$ is the
highest critical speed in vacuum, infinite speed and reversal of the arrow 
of time occur only in frames moving with respect to the vacuum rest frame
at speed $V~\geq ~c^2C^{-1}$ . Finite critical speeds
of superluminal sectors, as measured by ordinary matter in frames
moving at $V~\neq ~0$ , are anisotropic. Therefore, directional detection 
of superluminal particles would allow to directly identify the effective vacuum 
rest frame for each superluminal sector.
\vskip 5mm
{\bf 4b. Non-covariance of dynamics}
\vskip 5mm
Interaction between particles from
different dynamical sectors is, for simple physical
reasons, expected to depend on the rest frame of the system.
For instance, Lorentz contraction has an intrinsic physical meaning in
the vacuum rest frame and, at equal speed, is different for particles
from different dynamical sectors. Therefore, the relative size of two such
particles moving
with the same speed will depend on their motion with respect to the absolute
rest frame. This should in principle influence their interaction properties.
\vskip 4mm
Assuming that the above particles can be dealt with as spherical 
extended objects
of radius $r_i$ , and neglecting spin, we can as an example
attribute to particle $i$ ,
with position  ${\vec {\mathbf x_i}}$ and 
moving at speed ${\vec {\mathbf v_i}}$ ,
a form factor depending on ${\vec {\mathbf x}}~-~{\vec {\mathbf x_i}}$:
\equation
F_i~({\vec {\mathbf x}}~,~{\vec {\mathbf x_i}}~,t)~~=
~~f_i~[({\vec {\mathbf x_L}}~-~
{\vec {\mathbf x_i}}_{\mathbf ,L}~-
~{\vec {\mathbf v_i}}~t)^2~r_i^{-2}~(1~-~v_i^2c_i^{-2})^{-1}~~+~~
({\vec {\mathbf x_\perp }}~-~
{\vec {\mathbf x_i}}_{\mathbf ,\perp })^2~r_i^{-2}]
\endequation
where ${\vec {\mathbf x_i}}_{\mathbf
{,L }}=v_i^{-2}
({\vec {\mathbf v_i}}.{\vec {\mathbf x_i}})~{\vec {\mathbf v_i}}$~ ,~
${\vec {\mathbf x_i}}_{\mathbf {,\perp }}=
{\vec {\mathbf x_i}}-{\vec {\mathbf
x_i}}_{\mathbf {,L }}$ and similarly for the longitudinal and transverse 
components of ${\vec {\mathbf x}}$ with respect to the direction of
${\vec {\mathbf v_i}}$ .
Taking as before ${\vec {\mathbf V}}~=~{\vec {\mathbf v_i}}$ ,
we find under an "ordinary" Lorentz transformation with relative speed 
${\vec {\mathbf V}}$ the transformed form factor 
$F'_i~({\vec {\mathbf x'}}~,~{\vec {\mathbf x'}}~,~t'$) given by:
\equation
F'_i~~=
~~f_i~[({\vec {\mathbf x'}}_{\mathbf L}~-
~{\vec {\mathbf x'_i}}_{\mathbf ,L})^2~r_i^{-2}~(1~-~V^2c_i^{-2})^{-1}
~(1~-~V^2c^{-2})~~+~~
({\vec {\mathbf x'}}_{\mathbf \perp }~-~
{\vec {\mathbf x'_i}}_{\mathbf ,\perp })^2~r_i^{-2}]
\endequation
Where ${\vec {\mathbf x'}}$ and ${\vec {\mathbf x'_i}}$ are the 
Lorentz-transformed coordinates and their longitudinal and transverse 
components with respect to the direction of  ${\vec {\mathbf V}}$
are defined in the same way as before.
The relative longitudinal size $l_i~l_0^{-1}$ , where $l_0$ and $l_i$ are
respectively the length  
of an ordinary 
particle and of particle $i$ taken to be at rest 
in the vacuum rest frame, turns under the above
Lorentz transformation into
$l_i~l_0^{-1}~(1~-~V^2c^{-2})^{-1/2}~(1~-~V^2c_i^{-2})^{1/2}$ 
in the new intertial frame if, in both frames, the length is measured in
the direction of  ${\vec {\mathbf V}}$ .
When $V~\rightarrow ~c$ , the longitudinal size of a superluminal particle,
as measured by ordinary matter, will tend to infinity:
the ordinary particles become infinitely thin in the 
longitudinal direction, as
compared to the superluminal ones. Dynamics should be sensitive to this
contraction, which reflects the interaction of moving particles with the
vacuum.
\vskip 4mm
To obtain a quantum lagrangian corresponding to the classical lagrangian
(9), we can write ${\vec {\mathbf p_i}}~=~(h/2\pi )~{\vec {\mathbf k_i}}$ ,
where $h$ is the Planck constant, and:
\equation
{\vec {\mathbf v_i}}~~=~~{\vec {\mathbf p_i}}~c_i~(p_i^2~+~m_i^2c_i^2)^{-1/2}
\endequation
from which, assuming for simplicity the particles to be spinless and neutral,
we can according to (10) build for particle $i$ the free lagrangian:
\equation
L_{i,free}~~=~~-~\int ~(k_i^0)^{-1}~d^3{\vec {\mathbf k_i}}~
{\mathbf a^{\dagger }_i}({\vec {\mathbf k_i}})
{\mathbf a_i}({\vec {\mathbf k_i}})
~m_i^2c_i^3~[(2\pi )^{-2}h^2k_i~^2+~m_i^2c_i^2]^{-1/2}
\endequation
where $k_i^0~=~[k_i~^2+~4\pi ^2~h^{-2}~m_i^2c_i^2]^{-1/2}~
=~2\pi ~p_i^0h^{-1} $ ,
${\mathbf a^{\dagger }_i}({\vec {\mathbf k_i}})$
creates a particle of type $i$ with
wave vector ${\vec
{\mathbf k_i}}$ , and ${\mathbf a}({\vec {\mathbf k_i}})$ 
annihilates such a particle. The normalization constraint for the operators is:
\equation
N_i~~=~~\int ~(k_i^0)^{-1}~d^3{\vec {\mathbf k_i}}~
{\mathbf a^{\dagger }_i}({\vec {\mathbf k_i}})
{\mathbf a_i}({\vec {\mathbf k_i}})
\endequation
where $N_i$ is the total number of particles of type $i$.
Assuming the above particles to be scalars, and following the standard 
construction of free quantum fields (e.g. [8]), 
the creation and annihilation operators can be written in terms of the 
the scalar field $\Phi _i ({\vec {\mathbf x}}~,~t)$:
\equation
{\mathbf a_i}^{\dagger}~({\vec {\mathbf k_i}})~~~=~~~(8\pi) ^{-3/2}~~
[{\mathbf \alpha _i}^{\dagger}~({\vec {\mathbf k_i}})~+~
{\mathbf \beta _i}^{\dagger}~({\vec {\mathbf k_i}})]
\endequation
where:
\equation
{\mathbf \alpha _i}^{\dagger}~({\vec {\mathbf k_i}})~~~=
~~~k_i^0~~\int~ d^3{\vec {\mathbf x}}~~exp~[i~({\vec {\mathbf k_i}}.
{\vec {\mathbf x}}~-~
k_i^0c_it)]~~\Phi _i({\vec {\mathbf x}}~,~t)
\endequation
\equation
{\mathbf \beta _i}^{\dagger}~({\vec {\mathbf k_i}})~~~=~~~
-~ic_i^{-1}~\int~ d^3{\vec {\mathbf x}}~~
exp~[i~({\vec {\mathbf k_i}}.{\vec {\mathbf x}}~-~
k_i^0c_it)]~~\partial /\partial t ~\Phi _i({\vec {\mathbf x}}~,~t)
\endequation
and similarly for ${\mathbf a_i}({\vec {\mathbf k_i}})$ . We are thus led to 
the free lagrangian:
\equation
L_{i,free}~~=~~\int~ d^3{\vec {\mathbf x}}~{\mathcal{L}}_i
\endequation
where:
\equation
{\mathcal{L}}_i~~=~~-~(4\pi)^{-1}~h~c_i~[m_i^2c_i^2~(h/2\pi )^{-2}~\Phi _i^2~
~-~c_i^{-2}~(\partial \Phi _i/\partial t)^2 ~+
~({\vec {\mathbf \nabla }} \Phi _i) ^2]
\endequation
and we can associate to (10) the quantum lagrangian:
\equation
L~~=~~\Sigma _{i=1}^N~~L_{i,free}~~+~~L_{int}
\endequation
where $L_{int}$ is the interaction lagrangian. Each sectorial free lagrangian 
density ${\mathcal{L}}_i $
is a scalar with respect to Lorentz transformations with $c_i$ as
the critical speed parameter, but not with respect to other Lorentz
transformations. In spite of the loss of universal Lorentz covariance, there 
seems to be
no obvious effect tending to spoil the consistency of the quantum field
theory involving superluminal particles (see also Subsection 4d). 
\vskip 5mm
{\bf 4c. Quantum wave equations}
\vskip 5mm 
In the previous subsection, we considered the situation where, in the vacuum
rest frame, all free quantum particles satisfy 
Klein-Gordon equations with $c_i$
as the critical speed parameter for particle $i$ . This was inspired by
our knowledge of "ordinary" particles. But
we can also address the following question: what is the most general local
wave equation for a particle in the Universe we just described? 
Assuming that the $SU(2)$ invariant 
equations of motion are linear in the wave function 
$\phi $ and can be formulated locally in the
absolute local
rest frame in terms of first and second-order derivatives, a free particle wave
function at time $t~=~t_0$ will satisfy in the vacuum rest frame the equation:
\equation
-~A~\partial^2 \phi /\partial t^2~~+~~iB\partial \phi /\partial t~~+~~\Sigma 
_{j=1}^3\partial ^2 \phi /\partial x_j^2~~-~~D~\phi ~~~=~~~0
\endequation
where $A$ , $B$ and $D$ are functions of $t$ 
and the $x_j$ are the real coordinates of
the position vector ${\vec {\mathbf x}}$ . As long as 
effects at cosmic scale 
can be ignored, 
we can to a first approximation
assume that $A$ , $B$ and $D$  have constant values
and neglect the large scale time evolution of frequencies and  
wave vectors due to the cosmological redshift.
In such a scenario, if 
$B$ and $D$ can be neglected, the cosmological redshift implies as in [7] that 
the emitted and observed
frequencies $\nu _e$ and $\nu _o$ 
of radiation will be related by the equation:
\equation
\nu _o ~a_o~~=~~\nu _e~a_e
\endequation
which also applies to proper rates of events (time dilation).
Thus, our scenario is close to the natural predictions of standard cosmology.
If $B$ and $D$ do not vanish (they give the mass and rest energy of the
particle), the dependence on cosmic time of $BA^{-1}$ and $DA^{-1}$
may be a nontrivial problem. $A$ can be set constant by local time rescaling,
which implicitly modifies the local time scale to describe the evolution
of the Universe.
\vskip 4mm
In the case (forbidden by Lorentz invariance) where 
$B~\neq ~0$ , equation (37) is not self-conjugate. Assuming, for simplicity,
that $\phi $ is a scalar, its complex conjugate $\phi ^*$ will satisfy 
in the vacuum rest frame the
wave equation:
\equation
-~A~\partial^2 \phi ^*/\partial t^2~~
-~~iB\partial \phi ^*/\partial t~~+~~\Sigma
_{j=1}^3\partial ^2 \phi ^*/\partial x_j^2~~-~~D~\phi ^*~~~=~~~0
\endequation   
In terms of energy and momentum, the solutions of equation (37) are:
\equation
E~=~(4\pi)^{-1} h~A^{-1}~[B~+~(B^2~+~4Ak^2~+~4AD)^{1/2}] 
~~~~~~~~~~~~~~~(solution~1) 
\endequation
\equation
E~=~(4\pi)^{-1} h~A^{-1}~[B~-~(B^2~+~4Ak^2~+~4AD)^{1/2}] 
~~~~~~~~~~~~~~~(solution~2)
\endequation 
whereas equation (39) admits the solutions:
\equation
E~=~(4\pi)^{-1} h~A^{-1}~[-B~+~(B^2~+~4Ak^2~+~4AD)^{1/2}]
~~~~~~~~~~~~~~~(solution~3)
\endequation
\equation
E~=~(4\pi)^{-1} h~A^{-1}~[-B~-~(B^2~+~4Ak^2~+~4AD)^{1/2}] 
~~~~~~~~~~~~~~~(solution~4)
\endequation
\vskip 4mm
\noindent
where, as usual, ${\vec {\mathbf p}}~=
~h~(2\pi)^{-1}~{\vec {\mathbf k}}$ and
${\vec {\mathbf k}}$ is the wave vector. The speed of the particle is:
\equation
{\vec {\mathbf v}}~~=~~{\vec {\mathbf \nabla }}_{\vec {\mathbf p}}~E
~~=~~{\vec {\mathbf p}}~{\mathcal{C}}^2~E_v^{-1}
\endequation
where $~E_v~=~{\mathcal{C}}~(p^2~+~m^2{\mathcal {C}}^2)^{1/2}$ , 
${\mathcal {C}}~=~A^{-1/2}$ , $m~=~(4\pi )^{-1}~h~(B^2~+~
4D{\mathcal {C}}^{-2})
^{1/2}$ 
and, solving the equations in  
${\vec {\mathbf p}}$ , we get the standard relativitic expression:
\equation
{\vec {\mathbf p}}~=~m~{\vec {\mathbf v}}~(1~-~v^2{\mathcal{C}}^{-2})^{-1/2}
\endequation
\par
Then, with respect to Lorentz transformations with critical speed parameter
$\mathcal{C}$ , the
energy is a linear combination of: a) a Lorentz scalar, given by the term 
$E_s~=~\pm (4\pi )^{-1}~h~BA^{-1}$ in (40~-~43);  
b) the time component of a four-vector, given 
in (40~-~43) by the term $E_v~=~\pm ~(4\pi )^{-1}~h~
A^{-1}~(B^2~+~4Ak^2~+~4AD)^{1/2}$ . 
The rest energy and the inertial mass times the square of the critical speed, 
which are identical in the case of standard relativity, are different
in the present case. Assuming $D~>~0$ , and
following standard procedures of field theory (e.g.
[8]), we may consider associating
to solutions 1 - 4 in (40 - 43) two 
scalar fields describing neutral particles:
a field $\Phi _1$
for solutions (1) and (4) and a field $\Phi _2$ for solutions (2) and (3). 
But none of such fields would satisfy a second-order wave equation like
(37) or (39). Instead, we may attempt to build a field
$\Phi $ associated to solutions
(1) and (2) and satisfying equation (37). 
Its conjugate $\Phi ^{\dagger }$ would
then correspond to solutions (3) and (4) and satisfy equation (39). 
In expressions of the same type as (30) and (32), 
the volume element $(k^0)^{-1}~d^3{\vec {\mathbf k}}$
should use instead of $k^0$ its four-vector component
$k^0_v~=~(k^2~+~m^2{\mathcal {C}}^2)^{1/2}$ 
ignoring the scalar term $k^0_s~=~\pm BA^{-1}/2$, whereas the exponential 
$exp~[i~({\vec {\mathbf k}}.{\vec {\mathbf x}}~-~
k^0{\mathcal {C}}t)]$ should use for $k^0$ the expression
$k^0~=~k^0_s~+~k^0_v$ .
This seems to be the right choice of quantum fields. In this case: a)
there would be no symmetry between positive and negative 
energy solutions inside each field; b) particle and antiparticle would have the
same inertial mass, but not the same rest energy.  
But it would be possible to derive
equations (37) and (39) from a free lagrangian density $\mathcal{L}$
in terms of $\Phi $ 
and $\Phi ^{\dagger }$ writing:
\equation
{\mathcal{L}}~~=~~-
~(4\pi)^{-1}~h~{\mathcal {C}}~({\mathcal {L}}_0~+~
i\rho ~{\mathcal {L}}_{\rho })
\endequation
where:
\equation
{\mathcal {L}}_0~~=~~m^2{\mathcal {C}}^2~(h/2\pi )^{-2}~\Phi ^{\dagger
}~\Phi
~-~{\mathcal {C}}^{-2}~\partial \Phi ^{\dagger} /
\partial t ~\partial \Phi \partial t~+
~{\vec {\mathbf \nabla }} \Phi ^{\dagger }.{\vec {\mathbf \nabla }} \Phi 
\endequation
\equation
{\mathcal {L}_{\rho}} ~~=~~
\Phi ^{\dagger}~\partial \Phi /\partial t~-~\partial \Phi 
^{\dagger } /\partial t ~\Phi 
\endequation 
$\rho ~=~B/2$ and ${\mathcal {L}}_\rho $ is basically a charge operator
accounting, in the lagrangian, 
for the difference in rest energy between the particle described by 
$\Phi $ and its antiparticle. Obviously, a term proportional to a charge in the 
lagrangian indicates the existence of a constant  potential (a
scalar with respect to space rotations, like the electric potential) 
in the vacuum rest frame. This in 
turn indicates that an effective charge has condensed in vacuum, and that this 
charge has locally the same motion as the vacuum rest frame.
\vskip 4mm
Similar considerations apply to fermions. Going back to Section 3 ,
we can assume that the spinor wave function $\mid \psi _n>$ is
locally approximated by a plane wave and, neglecting the cosmological redshift,
satisfies in the vacuum rest frame the equation: 
\equation
d/dt~\mid \psi _n>~=~-i~e_n~\mid \psi _n>
\endequation
which has solution:
\equation
\psi _n~({\vec {\mathbf x}}~,~t)~~=~~exp~[-ie_n(t - t _0)]~~\psi
_n~({\vec {\mathbf x}}~,~t _0)~~=
~~exp~[-ie_n(t -t _0)~+~i{\vec {\mathbf k}}.{\vec {\mathbf x}}]~~\psi _0
\endequation
with group 
velocity 
${\vec {\mathbf v}}~~\simeq ~4\pi~h^{-1}~
s~(e_n-e_{n-1})~t~{\vec {\mathbf k}}/k$~. 
If the $n$-dependence of
$e_n~n^{-1}$ can be neglected (massless case), 
the constant $2\pi h^{-1}~e_n k^{-1}$ 
is the critical speed of the particle in vacuum
and the helicity is equivalent to chirality. 
If we apply equations (37) and (39) to the spinor wave function, we get
similar solutions for energy in terms of momentum. But we can also linearize
the equations introducing the real constants 
$\alpha $ , $\delta $ and $\epsilon $ , and writing for (37):
\equation
(\alpha k_0~+~{\vec {\mathbf \sigma}}. {\vec {\mathbf k}}~+~\delta )~
\psi _1~~+~~\epsilon ~\psi _2~~=~~0
\endequation
\equation
(\alpha k_0~-~{\vec {\mathbf \sigma}}. {\vec {\mathbf k}}~+~\delta )~
\psi _2~~+~~\epsilon ~\psi _1~~=~~0
\endequation
which lead to:
\equation
[(\alpha k_0~+~\delta )^2~-~k^2~-~\epsilon ^2]~\psi _1~~=~~
[(\alpha k_0~+~\delta )^2~-~k^2~-~\epsilon ^2]~\psi _2~~=~~0
\endequation
and, for (39):
\equation
(\alpha k_0~+~ {\vec {\mathbf \sigma}}. {\vec {\mathbf k}}~-~\delta )~
\psi _1^*~~+~~\epsilon ~\psi _2^*~~=~~0 
\endequation
\equation
(\alpha k_0~-~ {\vec {\mathbf \sigma}}. {\vec {\mathbf k}}~-~\delta )~
\psi _2^*~~+~~\epsilon ~\psi _1^*~~=~~0
\endequation
leading to:
\equation
[(\alpha k_0~-~\delta )^2~-~k^2~-~\epsilon ^2]~\psi _1^*~~=~~
[(\alpha k_0~-~\delta )^2~-~k^2~-~\epsilon ^2]~\psi _2^*~~=~~0
\endequation
\par
We are thus led to a pair of complex-conjugate generalized Dirac equations
where, in the chiral representation, a constant term has been added to 
the Dirac operator.
\vskip 4mm
The above discussion points at a possible intrinsic breaking of Lorentz 
invariance, and of the symmetry between particles and antiparticles,
that can be explored experimentally even if the effect is very
small. If such a phenomenon happens, and 
even if it occurs only in a superluminal
sector, it can naturally be at the origin of the asymmetry between matter
and antimatter in the Universe.
A relevant question is how universal would be (if it exists) 
the value of $B$ inside a given
sector. Experiment seems to indicate that $A$ is universal inside the
ordinary sector, whereas the effective value of
$D$ varies considerably for reasons that may be
due to spontaneous symmetry breaking.
Lacking a more detailed dynamical description of the phenomenon, 
we leave this question open.
\vskip 5mm
{\bf 4d. Causality, spin and statistics}
\vskip 5mm
If the number of sectors is finite and all
critical speeds are finite, causality
is not violated but adapted to the existence of several
critical speeds in vacuum. It holds explicitly in the vacuum rest frame,
where no signal can propagate faster than the highest critical speed $C$ .
If the interaction between sectors with different critical speed
in vacuum is weak, 
we expect by continuity the usual spin-statistics connection
to hold, as it holds in the limit
where different sectors would not interact. 
Then, the standard arguments for the spin-statistics
connection [9] remain valid with similar assumptions 
generalized to the new situation.
As stressed by many authors (e.g. [10]), causality is the only known 
principle allowing nowadays to demonstrate the observed relation between 
spin and statistics.
\vskip 4mm
Even with the breaking of Lorentz invariances due to interaction between 
different sectors, the existence of a maximum critical speed in the vacuum
rest frame and the causality condition with respect to $C$ 
in this frame, are enough to
enforce the standard connexion between spin and statistics
for all sectors (the ordinary sector and the superluminal ones). For 
instance, as discussed 
in [10], if the canonical quantization for bosons is applied to fermions,
the commutator between a 
free Dirac fermion field at $(t~,~ {\vec {\mathbf x}})$
and its hermitic conjugate at $(t'~,~ {\vec {\mathbf x'}})$ is given by a
Dirac operator acting on the integral:
\equation
\Delta _{1,i} (t~,~ {\vec {\mathbf x}}~;~t'~,~ {\vec {\mathbf x'}})~~=~~
(2\pi )^{-3}~\int ~(2k_0)^{-1}~d^3{\vec {\mathbf k}}~cos [k_0~c_i~(t-t')~-~
{\vec {\mathbf k}}.({\vec {\mathbf x}}~-~{\vec {\mathbf x'}})]
\endequation
where it can be seen that the function 
$\Delta _{1,i} (t~,~ {\vec {\mathbf x}}~;~t'~,~ {\vec {\mathbf x'}})$  and 
its derivatives do not vanish at $t~=~t', 
{\vec {\mathbf x}}~\neq~{\vec {\mathbf x'}}$ . Therefore, the incompatibility 
between causality and the "wrong" statistics remains. Similarly, the
compatibility between the "right" statistics and causality survives if
causality is understood as being valid in the vacuum rest frame with respect
to the highest critical speed $C$ , as common sense suggests.
\vskip 6mm
{\bf 5. SOME EXPERIMENTAL CONSIDERATIONS}
\vskip 5mm
For obvious reasons, the above analysis (which is far from exhausting the
basic theoretical problems) was necessary before considering any possible
experimental search for superluminal particles.
Experimental implications of the present study
will be discussed in detail in a forthcoming paper
but, from 
the above discussion, it clearly
appears that they are important and that
several effects of superluminal sectors
may be measurable in future experiments. Contrary to tachyons, which are 
defined and studied within the original framework of standard relativity 
[11] considered an absolute property
of space-time, the superluminal particles we propose imply by themselves the
abandon of this principle
(although the space-time felt by our laboratory
remains the "ordinary" minkowskian space-time). 
The new particles have specific experimental signatures,
different from those of tachyons [12]. 
However, since they can travel much faster than
light, they can indeed produce some "space-like" phenomena,
in astrophysics as well as at accelerators:
\vskip 4mm
- A superluminal particle moving at speed ${\vec {\mathbf v}}$ with 
respect to the vacuum rest frame, 
and emitted by an astrophysical object, can reach an observer, 
moving with laboratory
speed  ${\vec {\mathbf V}}$ with respect to the same frame, at a time (as 
measured by the observer) previous to the emission time. Such a phenomenon will
happen if  ${\vec {\mathbf v}}.{\vec {\mathbf V}}~>~c^2$ , and the 
emitted particle will be seen to evolve backward in time (but it evolves 
forward in time in the vacuum rest frame). If they interact several times with
the detector,
superluminal particles can
be a directional probe preceding the detailed observation of
astrophysical phenomena, such as explosions releasing
simultaneously neutrinos, photons and superluminal particles.
Directional detection of high-speed
superluminal particles in a large underground or 
underwater detector would allow
to trigger a dedicated astrophysical observation in the direction of the sky 
determined by the velocity of the superluminal particle(s).
If $d$ is the distance between the observer and the astrophysical object,
and $\Delta t$ the time delay between the detection of the superluminal
particle(s) and that of photons and neutrinos, 
we have: $d~\simeq ~c\Delta t$ .
\vskip 4mm
- Although we expect coupling constants to be small, 
superluminal particles evolving backward in time can be produced
at high-energy accelerators if the energy range $E~\stackrel{>}{\sim }~
m_i~c_i^2$ is reached for some of such particles. 
Large detectors with high time resolution may in
some cases be able to use timing to identify the production of superluminal
particles, especially if they are produced by pairs. If, as suggested 
by standard cosmology considerations, a typical speed for 
superluminal particles in our laboratory rest frame is $v ~\approx
~10^3~c~\approx ~10^{11}~ms^{-1}$ , a detector 
of radius $\approx 10~m$ with time
resolution $\approx 10^{-9}s$ (not incompatible with the LHC program)
would in principle
be able to distinguish between ordinary
and superluminal particles, at least in some of the relevant events. Larger and 
faster detectors would be necessary at a later stage in order 
to measure a high superluminal speed or to
identify a particle evolving backward in
time, assuming that it interacts sufficiently with the detector
(e.g. through "Cherenkov" effect in vacuum). 
\vskip 4mm
A neat experimental distinction can be set between the new 
superluminal particles 
and the "bradyons" and "tachyons" of ref. [12] . Even if some experimental
predictions show analogies between our particles and the tachyons,
the main properties of the new particles are closer to those of bradyons
(the ordinary subluminal particles of standard particle physics). 
The new superluminal particles are
actually "superbradyons", i.e. bradyons with a critical speed  
in vacuum higher than $c$ . Like bradyons, they have positive mass
as well as positive energy in the laboratory rest frame. 
Contrary to tachyons [13] , they can emit
"Cherenkov" radiation in vacuum [1~-~5] and, although tachyon theory also hints
to a vacuum rest frame [14], the physical reason is fundamentally different:
there is no need for any "reinterpretation principle" in the case of our 
superluminal particles.   
\vskip 4mm
The analogy with sine-Gordon solitons (see [1~-~5]) can help to understand
the difference between our "superbradyons" and the "tachyons" of previous
theories. In a galilean space-time, we can build dynamical systems 
satisfying equations of dalembertian type (i.e. with an invariance identical
to Lorentz invariance, but with a different critical speed related to
the properties of the dynamical system). Solitons of such systems, 
if they form a closed self-interacting system, would feel a "relativistic"
space-time with their own critical speed playing the role of the speed of
light. Such a "micro-universe" would, in some sense, fake our Universe  
in a finite region of our time. Since the dynamical system has a rest frame,
a "vacuum rest frame" is automatically set (like in our Universe, if particles
are solitons or topological singularities of vacuum, and if vacuum has a 
natural rest frame set by cosmology) "Lorentz" contraction 
in the "micro-universe" is not relative, but absolute, even
if a serious but isolated
observer made of solitons would be led to formulate a relativity 
principle and believe erroneously that such a contraction is indeed relative.
However, the dalembertian equation is in general a continuum approximation
to a discrete system, or the "large distance" limit of a more complex 
dynamics. It is not valid below a certain
distance scale $\kappa ^{-1}$ , where $\kappa $ is a critical wave vector
scale of the system. 
\vskip 4mm
The above "micro-universe" is not tachyonic, but bradyonic: 
solitons have always positive 
mass and energy.
As long as, even with "Lorentz" 
contraction, the longitudinal size of the
soliton remains much larger than $\kappa ^{-1}$ , "Lorentz" invariance
guarantees that a soliton, if stable when at rest in the absolute rest frame
(the rest frame of the dynamical system),
remains stable when it is accelerated to a nonzero speed with respect to
the same frame.
However, if the soliton is accelerated, with respect to the
absolute rest frame, to a speed such that by Lorentz contraction its 
longitudinal size
becomes $\stackrel{<}{\sim }~\kappa ^{-1}$ , the "relativity principle" felt by 
the soliton micro-universe does no longer apply. Then,
the previously stable
soliton can decay into other excitations, like unstable waves, 
and disperse its energy. A similar discussion can be raised for elementary
particles such as quarks and leptons, or their constituents:
can we break these particles, just by accelerating them to an inverse distance 
scale such
that relativistic kinematics does no longer hold and the structure of
vacuum manifests itself at a deeper level? The existence of very high-energy
cosmic rays seems to indicate that such a phenomenon can only
happen at energies beyond the reach of present and planned accelerators. 
However, the subject deserves attention from a long-term point of view,
not only for future very high-energy machines but also in the analysis of the
highest-energy cosmic rays. In a sectorial grand-unified theory with
spontaneous symmetry breaking, it may happen that the estimated mass of
a Higgs or intermediate boson be higher than 
${\boldmath e} _i~c_i^{-2}~\approx ~h~\kappa_i~c_i^{-1}$ , 
where $\kappa _i$
is the value of $\kappa $ for the sector under consideration
and ${\boldmath e}_i$ is a critical energy scale.  
It is possible that such a boson will never be formed, and similarly for
fermions undergoing the same phenomenon. 
\vskip 4mm
The question of whether the constant $B$ (related to some static 
charge condensed 
in vacuum) vanishes or not, arises already in
the ordinary sector. It could happen
that its existence be hidden by inertial masses 
(typically, if $AD~\gg ~B^2$).
Therefore, it seems sensible to attempt experiments with the so-called 
"massless" particles (photon and neutrinos) in order to possibly show the
effect. If $B~=~0$ , a sectorial Lorentz invariance is dynamically generated.
A possible evidence for $B~\neq ~0$ would be an asymmetry between 
neutrino and antineutrino oscillations. The cosmological role of the condensed 
charge, and of the
"anomalous" neutrino rest energy, would deserve further investigation. 
\vskip 4mm
It turns out from (18), (23) and (24) that, for a high-speed 
superluminal cosmic ray with critical speed $c_i~\gg ~c$ , 
the momentum, as measured in the laboratory,
does not provide directional 
information on the source, but on the vacuum rest frame. 
Velocity provides directional information on the source, 
but can be measured only if the 
particle interacts several times with the detector, which is far from
guaranteed, or if the superluminal particle is associated to a collective
phenomenon emitting also photons or neutrinos simultaneously.
>From (18), if in
the vacuum rest frame the detected particle has speed $v~\simeq ~c_i$ and
energy $E~\simeq ~p~c_i$ , the momentum ${\vec {\mathbf p'}}$ in the laboratory
rest frame will be dominated by the term $-~E~c^{-2}~{\vec {\mathbf V}}
~\simeq ~p~c_i~c^{-2}~{\vec {\mathbf V}}$ 
if $c_i~v~\gg ~c^2$ , while the velocity 
${\vec {\mathbf v'}}$ is dominated by the term ${\vec {\mathbf v}}~
(1~-~{\vec {\mathbf v}}.{\vec {\mathbf V}}~c^{-2})^{-1}$ ,
and the particle will be seen (in the laboratory) to evolve backward in time
if ${\vec {\mathbf v}}.{\vec {\mathbf V}}~>~c^2$ .
Again, standard cosmology suggests that the main speed range for cosmic 
superluminal particles, as measured on earth by a detector made of ordinary
matter, should be $v~\approx ~10^3~c$ . This would by itself be a signature.
As more schematically emphasized in [5] , one 
has $E'~\gg ~p'~c$ , which leads to
"back-to-back" events if the superluminal particle transfers most of its energy
to a pair of ordinary particles.
\vskip 4mm
In this paper, we have assumed that there exists an absolute rest frame of
vacuum, which: a) is the same for all superluminal sectors ; b) provides the
local rest frame for the expanding Universe; c) is close to the
local rest frame suggested by the study of cosmic background radiation. If
superluminal particles are found, it will be possible to directly check the
basic hypothesis of this simple scenario. It may happen that 
the expansion of vacuum has followed different
evolutions for the degrees of freedom linked to different sectors
or, for instance,
that the Universe is not really isotropic because of relative rotational
modes between these sets of degrees of freedom
(this anisotropy could even be spontaneously generated). 
In such case, it is likely
that a 
single (local) vacuum rest frame would not exist. The
experimental discovery of
particles belonging to different superluminal sectors 
would then 
(assuming the speed in the laboratory rest frame to be measurable) 
allow to
extract spectra of ${\vec {\mathbf v}}$ and ${\vec {\mathbf p}}$
(see Subsection 4b), perhaps
not necessarily in the range $v~\approx ~10^3~c$ or 
with a different directional dependence, for
each dynamical sector. It would then,
in any case, be possible for the first time to get a direct
insight into the inner structure of vacuum
(evidence for a non-empty vacuum has been provided 
by particle physics, e.g. through the discovery of the 
$W^{\pm}$ and $Z^0$). Ultimately, experiments 
at very high-energy accelerators or an observatory of superluminal cosmic
rays may be able to determine the value of $\kappa _i$ for
each superluminal sector by looking at the energy, momentum 
and velocity distribution
of the events. However, although the basic phenomenon can perhaps be neatly
observed,
the dynamical interpretation of the data is likely to be far from trivial.
\vskip 1cm
{\bf References}
\vskip 5mm
\par
\noindent
[1] L. Gonzalez-Mestres, "Properties of a possible class 
of particles able to travel faster 
than light", Proceedings of the Moriond Workshop on "Dark Matter in Cosmology,
Clocks and Tests of Fundamental Laws", Villars (Switzerland), January 21-28
1995 , Ed. Fronti\`eres, France. Paper astro-ph/9505117 of electronic library.
\par
\noindent
[2] L. Gonzalez-Mestres, "Cosmological implications of a possible class of
particles able to travel faster than light", Proceedings of the Fourth 
International Workshop on Theoretical and Experimental Aspects of 
Underground Physics, Toledo (Spain) 17-21 September 1995~, Nuclear Physics B
(Proc. Suppl.) 48 (1996). Paper astro-ph/9601090 .
\par
\noindent
[3] L. Gonzalez-Mestres, "Superluminal matter and high-energy cosmic rays",
May 1996 . Paper astro-ph/9606054 .
\par
\noindent
[4] L. Gonzalez-Mestres, "Physics, cosmology and experimental signatures of 
a possible new class of superluminal particles", to be published in the
Proceedings of the International Workshop on the Identification of Dark
Matter, Sheffield (England, United Kingdom), September 1996 . Paper
astro-ph/9610089 .
\par
\noindent
[5] L. Gonzalez-Mestres, "Physical and cosmological implications of a possible
class of particles able to travel faster than light", contribution to the
28$^{th}$ International Conference on High-Energy Physics, Warsaw July 1996 .
Paper hep-ph/9610474 .
\par
\noindent
[6] See, for instance, L. Bergstrom et al., "THE AMANDA EXPERIMENT: status and
prospects for indirect Dark Matter detection", same Proceedings as for ref.
[4]. Paper astro-ph/9612122 . 
\par
\noindent
[7] See, for instance, P.J.E. Peebles, "Principles of Physical Cosmology",
Princeton Series in Physics, Princeton University Press 1993 .  
\par
\noindent
[8] S.S. Schweber, "An Introduction to Relativistic Quantum Field Theory",
Row, Peterson and Company 1961 .
\par
\noindent
[9] See, for instance,
R.F. Streater and A.S. Wightman, "$PCT$, Spin and Statistics, and All
That", Benjamin, New York 1964 ; R. Jost, "The General Theory of
Quantized Fields", AMS, Providence 1965 .
\par
\noindent
[10]  C. Itzykson and J.B. Zuber, "Quantum Field Theory", McGraw-Hill 1985 . 
\par
\noindent
[11] The relativity principle was formulated by
H. Poincar\'e, Speech at the St. Louis International Exposition of 1904 ,
The Monist 15 , 1 (1905).
\par
\noindent
[12] See, for instance,
"Tachyons, Monopoles and Related Topics", Ed. by E. Recami,
North-Holland 1978 , and references therein.
\par
\noindent
[13] See, for instance, E. Recami in [10] .
\par
\noindent
[14] See, for instance, E.C.G. Sudarshan in [10] .
\end{document}